%% file: paperplb.tex
\documentclass{elsart}
\usepackage{epsfig}
\usepackage{rotate}
\usepackage{amssymb}
\usepackage{amsmath}

\newcommand\pair{e^+ e^-}

\newcommand\nudecay{\nu_4 \rightarrow \nu_\tau e^+  e^-}

\newcommand\nus{\nu_4}

\newcommand\C{\mathcal C}
\newcommand\ee{e^+e^-}

\begin{document}
\begin{frontmatter}
\title{\boldmath Search for heavy neutrinos\\
 mixing with tau neutrinos}
%\collab{NOMAD Collaboration}
\vspace{-0.cm}
\centerline{\bf NOMAD Collaboration}
\author[Paris]             {P.~Astier}
\author[CERN]              {D.~Autiero}
\author[Saclay]            {A.~Baldisseri}
\author[Padova]            {M.~Baldo-Ceolin}
%\author[CERN]              {G.~Ballocchi}
\author[Paris]             {M.~Banner}
\author[LAPP]              {G.~Bassompierre}
\author[Lausanne]          {K.~Benslama}
\author[Saclay]            {N.~Besson}
\author[CERN,Lausanne]     {I.~Bird}
\author[Johns Hopkins]     {B.~Blumenfeld}
% ...

\address[LAPP]           {LAPP, Annecy, France}
\address[Johns Hopkins]  {Johns Hopkins Univ., Baltimore, MD, USA}
% ...
\address[CERN]           {CERN, Geneva, Switzerland}
\address[Lausanne]       {University of Lausanne, Lausanne, Switzerland}
% ...
\address[Padova]         {Univ. of Padova and INFN, Padova, Italy}
\address[Paris]          {LPNHE, Univ. of Paris VI and VII, Paris, France}
% ...
\address[Saclay]         {DAPNIA, CEA Saclay, France}
%\address[ANSTO]          {ANSTO Sydney, Menai, Australia}
% ...

%\thanks[Perugia]         {Now at Univ. of Perugia and INFN, Perugia, Italy}

\clearpage
\begin{abstract}
We report on a search for heavy neutrinos ($\nus$) produced in
the decay $D_s\to \tau \nus$ at the SPS proton target followed by the 
decay $\nudecay$ in the NOMAD detector.\ Both decays are expected to occur
if $\nus$ is a component of $\nu_{\tau}$.\ 
 From the analysis of the data collected during the 1996-1998 runs
 with $4.1\times10^{19}$ protons on target, a single
 candidate event consistent with background expectations was found.\
%No evidence for this decay chain has been found 
This allows to derive  an upper limit on the mixing strength 
between the heavy neutrino and the tau neutrino  
in the $\nus$ mass range from 10 to 190 $\rm MeV$.\ Windows between the SN1987a 
and  Big Bang Nucleosynthesis lower limits and our result are 
still open for 
future experimental searches.\  The results obtained are used to 
constrain an interpretation of the time anomaly observed in the KARMEN1
detector.\
\end{abstract}
\begin{keyword} 
neutrino mixing, neutrino decay
\end{keyword}
\end{frontmatter}

\input motivation

\input nomad

\input method

\input analysis

\input background

\input result_p

\input karmen1

\input conclusion

\input bibliography
\newpage
  \begin{figure}
 \begin{center}
   \mbox{\hspace{-1.8cm}\epsfig{file=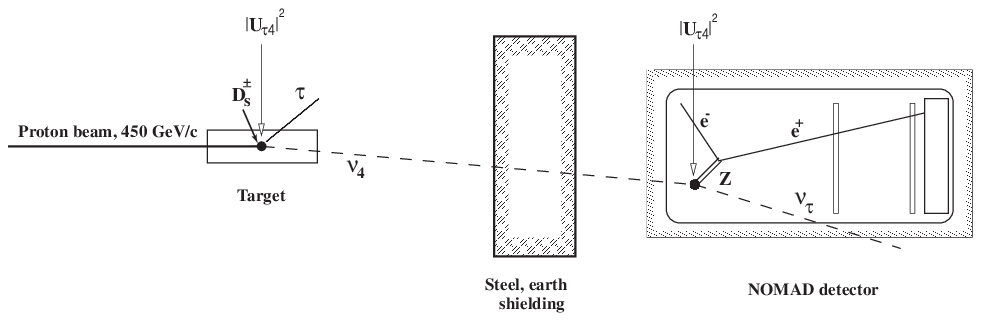,width=120mm}}
\end{center}
    \centering
\vspace{0.5cm}
    \caption{\em Schematic illustration of the production of $\nus$ and 
its detection via the $\nudecay$ decay.\ Because of its short lifetime
the $D_s$ decays immediately at the production point.}
%  \label{figure 1:}
\end{figure}

  \begin{figure}
.\vspace{-3.5cm}
   \mbox{\hspace{3.0 cm}\epsfig{file=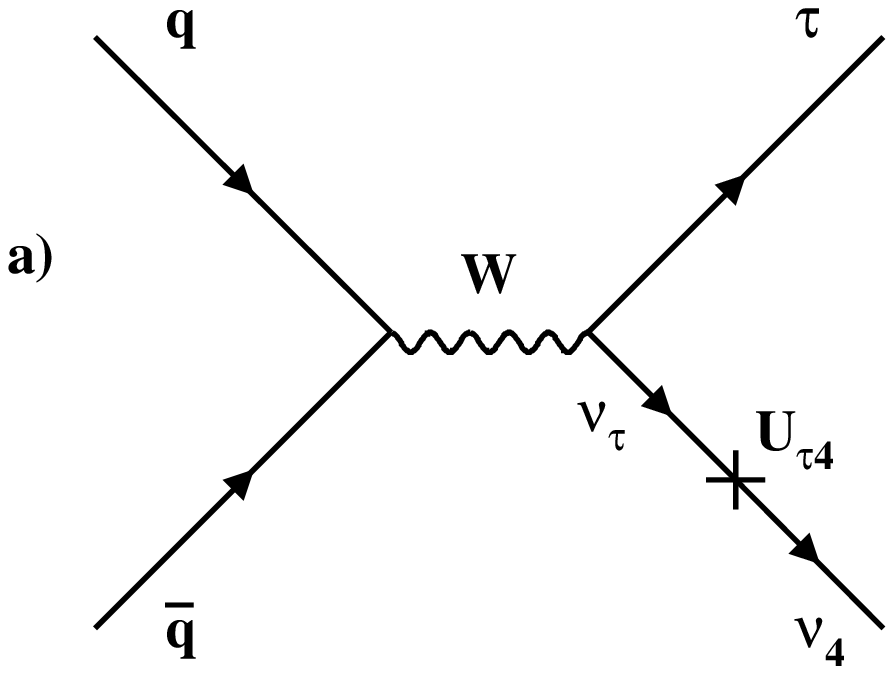,height=120mm}}
.\vspace{-6.5cm}
 \mbox{\hspace{3.0 cm}\epsfig{file=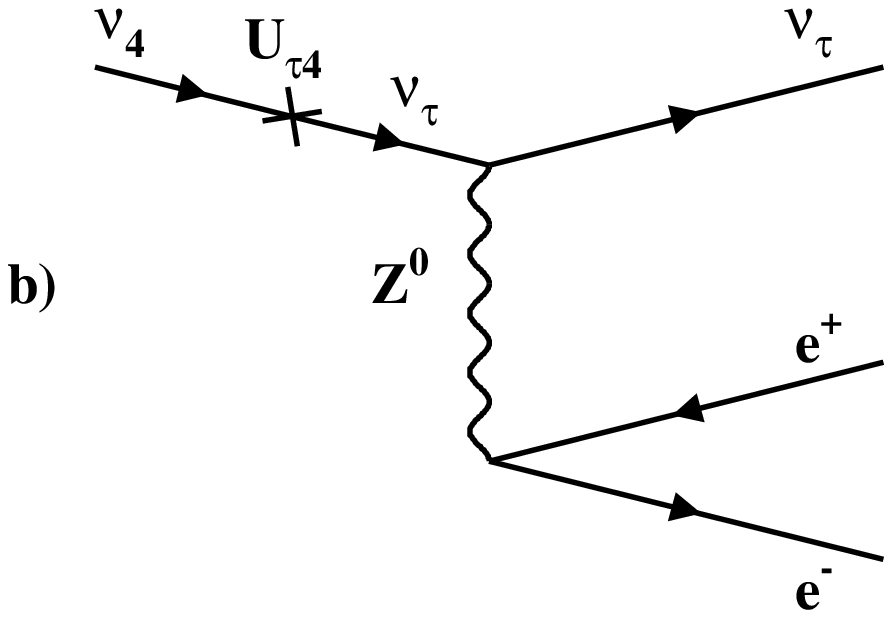,height=120mm}}
    \centering
    \caption{\em Feynman diagrams illustrating a) $\nus$ production  from 
 $D_s(\overline{D}_s)$ decay  and b) the decay of an isosinglet neutrino
$\nus$.}
  \label{figure 1:}
\end{figure}

\newpage
  \begin{figure}
 \begin{center}
   \mbox{\epsfig{file=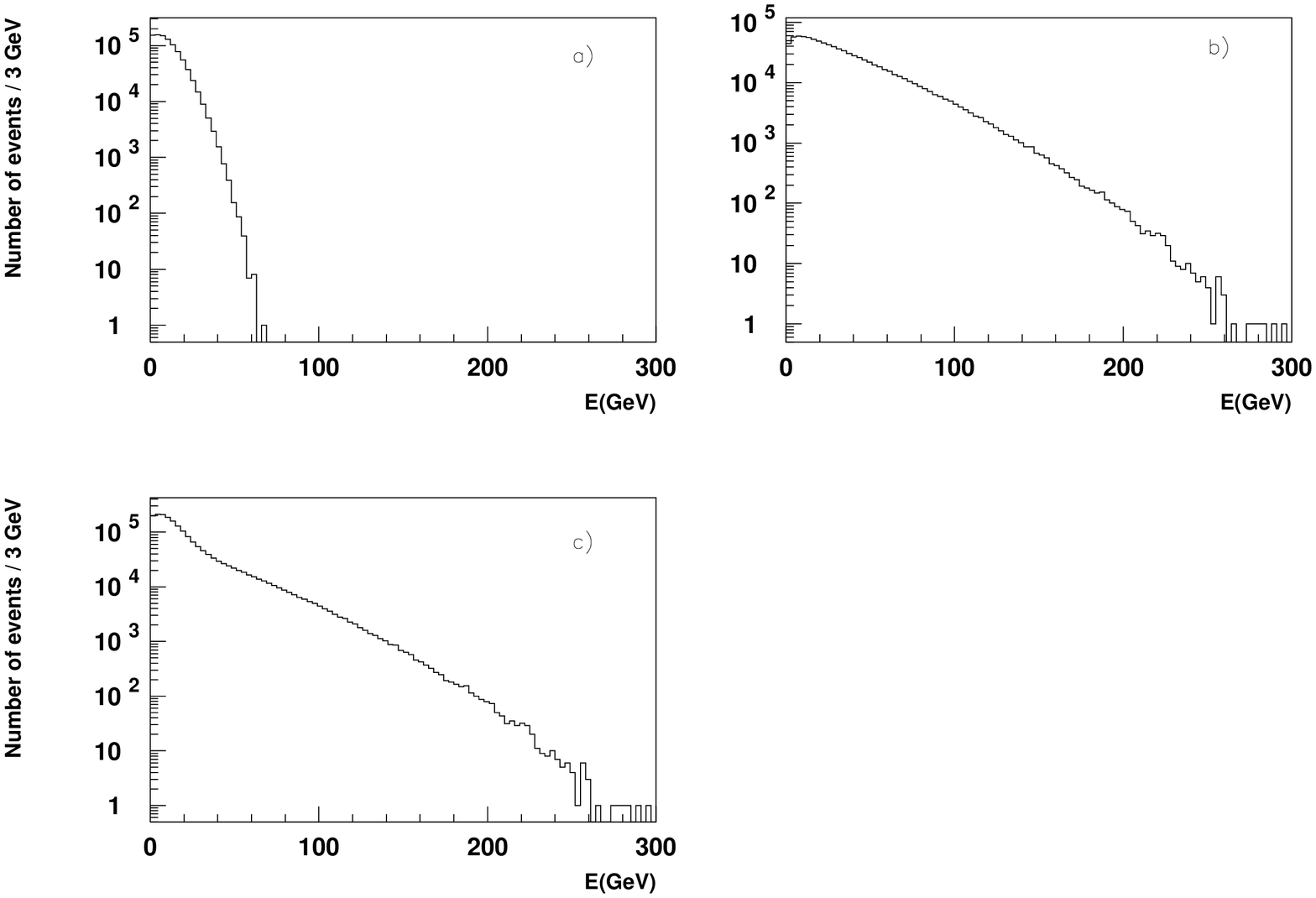, height=160mm,width=160mm}}
\vspace{0.5cm}
    \centering
    \caption{\em Combined energy spectra of tau neutrinos 
 pointing towards the NOMAD fiducial area from the 
proton target and from the beam dump a) originating directly from  $D_s(\overline{D}_s)$ decays;
b) originating 
from the decays of $\tau^{\pm}$ produced in $D_s(\overline{D}_s)$ decays; 
c) the sum of the two.\ The spectra 
are calculated for $10^{11}$ protons on target ({\em pot}).}
  \end{center}
  \label{figure 1:}
\end{figure}

\newpage
  \begin{figure}
 \begin{center}
   \mbox{\hspace{-1.5cm}\epsfig{file=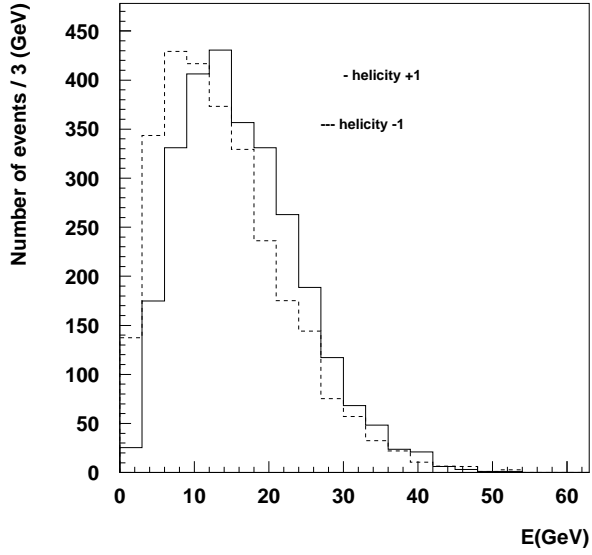,height=80mm}}
\end{center}
    \centering
    \caption{\em Energy distribution  of $e^+e^-$ pairs
 from  decays of a $\nus$ with 33.9 $\rm MeV$  mass 
for different $\nu_4$ helicities. The average $e^+e^-$ pair energies are 
 $<E>_{+1}=16.4~\rm GeV$ and $<E>_{-1}=14.1~\rm GeV$ for helicities +1 and -1,
respectively.} 
  \label{figure 1:}
\end{figure}

  \begin{figure}
 \begin{center}
   \mbox{\epsfig{file=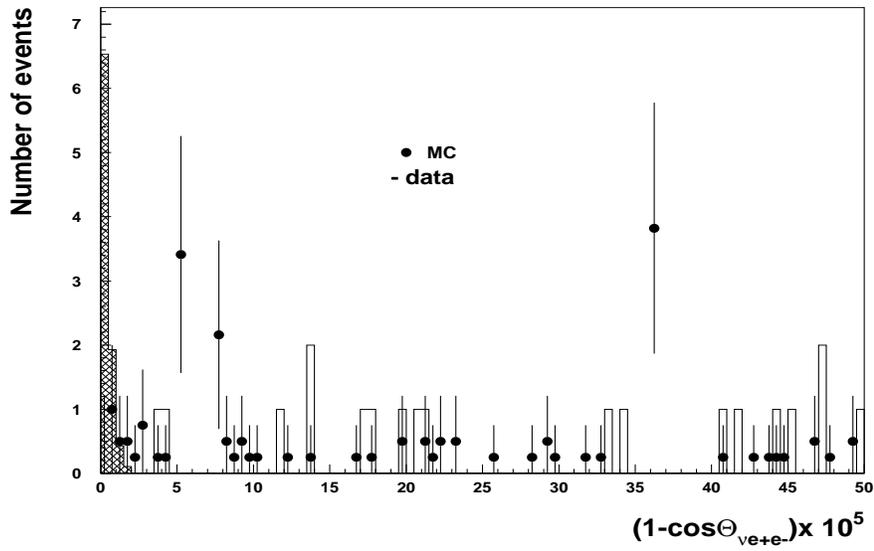,height=80mm,width=120mm}}
    \centering
    \caption{\em The final (1 - $cos\Theta_{\nu e^+ e^-)}$ distribution  
for the data and MC.\ The  dashed histogram represents the distribution 
expected from signal events.}
\end{center}
  \label{figure 1:}
\end{figure}

\newpage
  \begin{figure}
 \begin{center}
   \mbox{\epsfig{file=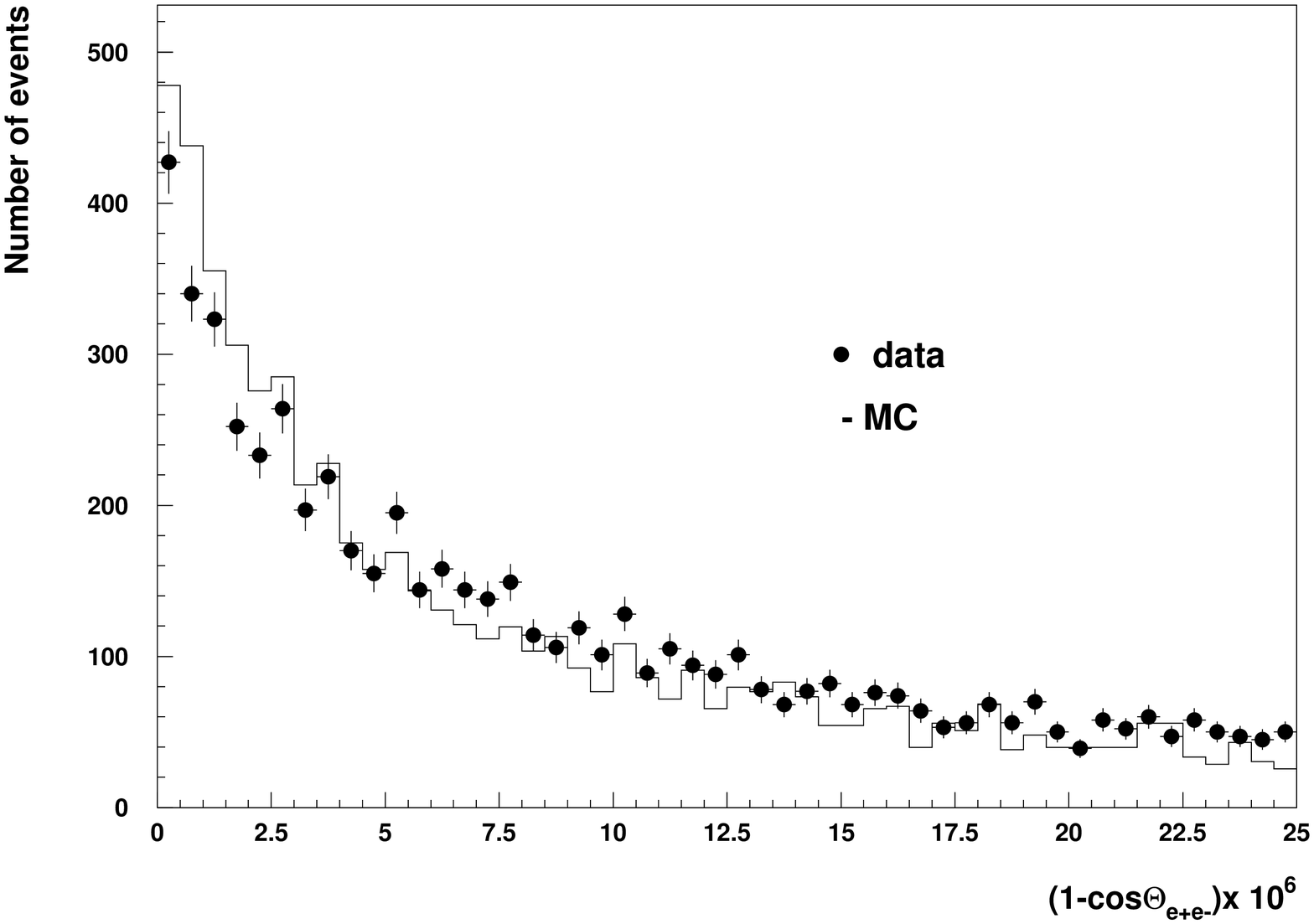,height=80mm,width=120mm}}
    \centering
    \caption{\em The (1 - $cos\Theta_{e^+ e^-}$) distribution  
for $\pair$ pairs from photons converted in the DC target at large distances 
from the primary vertex for the data and MC.}
\end{center}
  \label{figure 1:}
\end{figure}

  \begin{figure}
 \begin{center}
   \mbox{\epsfig{file=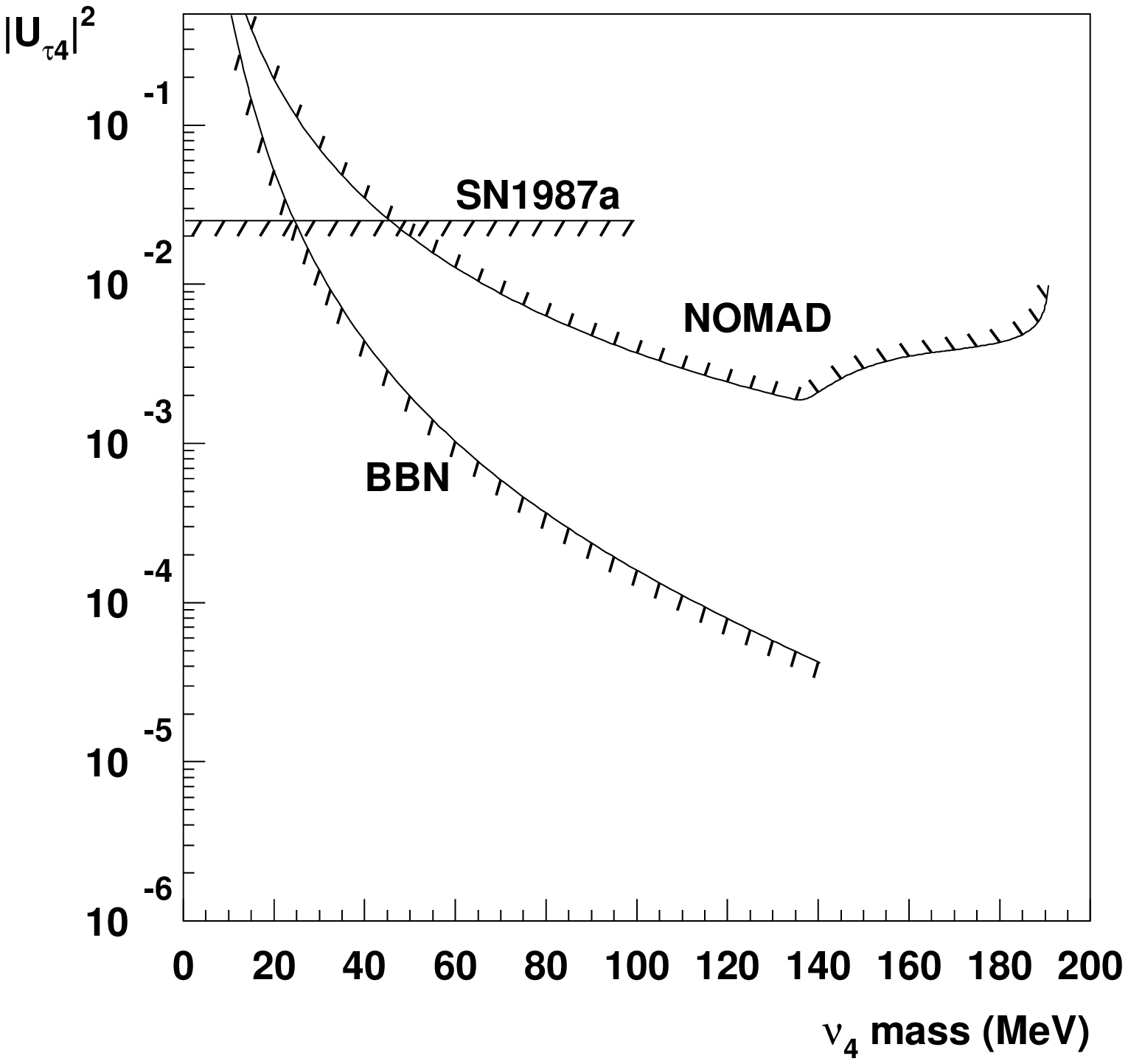,height=90mm,
width=120mm}}
\end{center}
%\vspace{2.cm}
    \centering
    \caption{\em  The NOMAD  90\% $CL$ upper limit, and the 
SN1987a and  Big Bang Nucleosynthesis lower limits 
 for the mixing strength 
$|U_{\tau 4}|^2$ as a function of the heavy neutrino mass.\ The SN1987a and 
BBN limits are reproduced from ref. \cite{dolgov1}. Note that  upper
limits of $|U_{\tau 4}|^2\lesssim 10^{-8}$ from SN1987a and 
$|U_{\tau 4}|^2\lesssim 10^{-10} \div 10^{-12}$ from BBN arguments have also been
derived for the corresponding mass ranges shown, respectively. 
For more details see ref. \cite{dolgov1}.}
  \label{figure 1:}
\end{figure}

\end{document}

%% file: motivation
\section{Introduction}
\label{sec:intro}

In the Standard Model all fundamental fermions have a right-handed component
that transforms as an isosinglet under the  
$SU(2)_L$ gauge group except  neutrinos, which are observed only
in left-handed form.\
 However, heavy  neutrinos which  
are decoupled from 
$W$ and $Z$ bosons and hence are mostly isosinglet (sterile) arise in many 
models that attempt to unify
the presently known interactions into a single gauge scheme, such as 
Grand Unified Theories or Superstrings inspired models \cite{theory}.\
They are also predicted in models trying to solve the 
problem of baryo- or leptogenesis in the Universe, in many extended 
electroweak models, such as left-right symmetric and  see-saw models 
\cite{theory}.\ Their masses are predicted to be within the $\rm GeV-TeV$ range.\
  The existence of a light ($\lesssim \rm eV$ or $\ll \rm eV)$ sterile neutrino is 
expected
in schemes  that attempt to solve the presently observed 
indication from atmospheric, solar and LSND experiments that neutrinos are 
massive, see e.g. \cite{ellis} and references therein.\
More generally one can also look for an isosinglet neutrino
 with intermediate mass such as in the $\rm keV-MeV$ range.\
For instance, such neutrinos with masses in the range 1 - 40 $\rm keV$ 
were recently considered as a candidate for warm dark matter \cite{dolgov3}.\  

 If heavy neutrinos exist, many crucial questions arise.\ 
For example, for massive  neutrinos the flavour eigenstates 
($\nu_e,~\nu_{\mu},~\nu_{\tau},...$) need not coincide with the mass 
eigenstates ($\nu_1,~\nu_2,~\nu_3,~\nu_4...$), but would, in general, 
be related
 through a unitary transformation.\ Such a generalised mixing:

\begin{equation}
\nu_l= \sum_i U_{li} \nu_i;~~~l=e,\mu,\tau,...,~i=1,2,3,4,...
\end{equation}

could result in neutrino oscillations when the mass differences are small, 
and in 
decays of heavy neutrinos when the mass differences are large.\ The relevant
questions are then, 
do heavy neutrinos mix with the ordinary neutrinos 
and, if so, what are the elements of the neutrino mixing matrix $U_{li}$?

Since there are no firm predictions for  $m_{\nus}$,
experimental searches of $\nus$ have been performed  
over a wide range of masses.\ A heavy
neutrino in the low mass region (below a few $\rm GeV$) 
has been searched for mainly in experiments on the leptonic decay of 
light mesons and in neutrino experiments \cite{heavynu} resulting in 
stringent upper limits on $|U_{li}|^2$ down to $10^{-7}$, thus
constraining the mixing amplitudes $|U_{e4}|^2$ and  $|U_{\mu 4}|^2$ 
for the electron and muon  neutrinos, respectively.\ Concerning the
 mixing strength $|U_{\tau 4}|^2$ for the MeV mass region studied in this 
experiment
a limit can be derived from some  earlier papers \cite{heavynu}, see also \cite{barger}.\ Note that stringent limits on $|U_{\tau 4}|^2$ 
 in this mass range have been recently  obtained from 
cosmological and astrophysical considerations \cite{dolgov1}.\
For  masses in the 3 GeV to  $\simeq200~\rm GeV$ range the LEP 
experiments have set limits on $|U_{\tau 4}|^2$ 
varying from $10^{-5}$ to $10^{-1}$ depending on the $\nus$ mass \cite{lep}.\
 
The motivation and purpose of this work is 
to search for a neutral heavy lepton $\nus$ which is dominantly
associated with the third generation of light neutrinos, $\nu_{\tau}$,
via the mixing term $|U_{\tau 4}|^2$.\ If such a particle exists it 
might be  produced in
the decay $D_s\to \tau \nus$ at the SPS proton target followed by the 
decay $\nudecay$ in the NOMAD detector as is illustrated in Figure 1 (see also Section 3).\
The experimental signature of 
these events is clean and they can be selected with small background
due to the excellent NOMAD capability for precise measurements of the 
$\ee$ pair direction.\
An additional motivation for the present study was the time anomaly
observed by the KARMEN1 experiment ( see Section 7).\ Although
in this paper we will assume that $\nus$ has  Dirac mass, 
the application of our results to the Majorana case is 
straightforward.\

%% file: nomad
\section{WANF and NOMAD detector}
\label{sec:detector}

The CERN West Area Neutrino Facility (WANF) beam line \cite{wanf} 
provides an essentially pure $\nu_{\mu}$
beam for neutrino experiments.\ 
It consists of a beryllium
target irradiated by  450 GeV  protons from the CERN SPS.\ 
The secondary hadrons of a given sign 
are focused with two magnetic elements, the horn and 
the reflector, located in front of a 290 m long evacuated decay 
tunnel.\ Protons that have not interacted in the target, secondary hadrons and 
muons that do not decay are absorbed by a 400 m thick shielding made of iron and 
earth.\   The NOMAD detector is located at 835 m from the neutrino target.\

 The detector is described in Ref.~\cite{nomad}.\ It consists of
a number of sub-detectors most of which are located inside a 0.4 T dipole 
magnet  with a  volume of 7.5$\times$3.5$\times$3.5 m$^{3}$:  
 an active target of 
drift chambers (DC)\cite{dc} with a mass of 2.7 tons
(mainly carbon),  an average density of $0.1~\rm g/cm^3$ and a total thickness
of about one radiation length ($\sim 1.0 X_0$)
 followed by 
a transition radiation detector (TRD)
\cite{trd}, a preshower detector (PRS), and an electromagnetic
calorimeter (ECAL).\ The PRS is composed of a plane of horizontal
and a plane of vertical proportional tubes
 preceded by a 9 mm thick lead plate ($1.6~ X_0$).\
% with fiducial mass of about 700 kg.\
 The ECAL consists of 875 lead-glass counters, each 
about 19 $X_0$ deep, arranged in a matrix of 
35 rows by 25 columns  \cite{ecal}.\  

 A hadron calorimeter (HCAL) and two
muon stations 
are located just after the magnet coils.\ 
A plane of scintillation counters,  $V$,  in front of the magnet was used to
veto  upstream neutrino interactions and muons incident on the detector.\ Two
 planes of scintillation counters $T_{1}$ and $T_{2}$ located before and after
the TRD were used for triggering purposes.\ The trigger for neutrino 
interactions or decays in the DC target was then $\overline{V} T_{1} T_{2}$.\

The electron identification efficiency in NOMAD is provided mainly 
by the TRD which has an efficiency of
more than 90\% for isolated electrons of momentum $1-50~\rm GeV/c$ for a 
charged pion  rejection factor greater than $10^3$ \cite{trd}.\

%% file: method
\section{Production and decay}

As follows from Eq. (1), any source of $\nu_{\tau}$   will 
produce all kinematically possible  massive eigenstates 
according to the appropriate mixings.\ In this search
the source of $\nu_{\tau}(\overline{\nu}_{\tau})$'s that can potentially 
generate $\nus$'s is $D_s(\overline{D}_s)$ mesons produced in the reaction
$p + Be\rightarrow D_s + X$  at 
the proton target and subsequently decaying leptonically: 
$D_s \to \nu_{\tau} + \tau$, $\tau \to \nu_{\tau} + X$ \cite{pdg}.\
Up to a $\nus$ mass of 190 MeV, the mass difference between the $D_s$ 
and the $\tau$, $\nus$'s can originate both from the $\nu_{\tau}$
produced directly in the $D_s$ decay and from the $\nu_{\tau}$ 
produced indirectly in the subsequent $\tau$ decay.\ For a $\nus$ mass 
larger than 190 MeV only $\nu_{\tau}$'s produced indirectly in  
$\tau$ decay can contribute.\ However these indirect $\nu_{\tau}$'s 
have a lower acceptance at NOMAD and a harder energy spectrum 
(Figure 3) resulting in a 
smaller probability to observe the decay $\nudecay$ in the NOMAD 
detector.\ Therefore this search is limited to $\nus$ masses smaller than 
190 MeV.\ If $\nus$ is a long-lived particle, the flux of $\nus$'s
would penetrate the downstream
shielding without significant attenuation  and would be observed in  NOMAD
through their $\nudecay$ decays as illustrated in Figure 1.\

 For neutrino masses below the $\pi^0$-meson mass, 
$m_{\nus}\lesssim m_{\pi}$, the dominant heavy neutrino visible
decay is $\nu \ee$  with a rate which,  
for small mixing,  is given by \cite{dolgov1,shrock} 

\begin{eqnarray}
\tau_4^{-1} \equiv \Gamma(\nu_4 \rightarrow \nu_{\tau} e^+ e^-) = K \big[ 
\frac{(1+g^2_L + g^2_R) G_F^2 m_{\nus}^5 |U_{\tau 4}|^2}{192 \pi^3} 
\big]
\end{eqnarray}

where $g_L = -1/2+ sin^2\theta_W$, $g_R=sin^2\theta_W$,  $K = 1(2)$ for Dirac(Majorana) 
particles.\  The corresponding diagram illustrating  
 the dominant contribution from neutral weak currents to this decay mode
is shown in Figure 2b.\ The branching ratio of the visible decay 
 is given by  

\begin{equation}
BR(\nus \rightarrow \nu_{\tau} e^+ e^-) = \frac{\Gamma(\nudecay)}{\Gamma_{tot}}
\simeq 0.14
\end{equation}

where the total rate $\Gamma_{tot}$ is dominated by 
the $\nus \to 3 \nu$ decay channel.\

For $m_{\nus}\gtrsim m_{\pi}$ the two-body decay channel   
$\nus \to \nu_{\tau} \pi^0$ opens up.\ This mode is 
phase space favoured and becomes dominant for $m_{\nus} \gtrsim 140~\rm MeV$
\cite{memo}.\

In this paper we have studied the decay $\nudecay$ 
as a possible manifestation of the presence of $\nus$'s 
in the neutrino beam.\ 
The occurrence of $\nudecay$ decays  would appear as an excess
of isolated $\pair$ pairs in NOMAD
above those expected from  standard neutrino
interactions.\ The decay $\nudecay$ cannot be distinguished 
from the anti-neutrino decay $\overline{\nu}_4\to \overline{\nu}_{\tau} 
e^+ e^-$ and the result of this search therefore referes to the sum of these 
two decays.\ 

The spectra of $\nu_{\tau}$'s produced in the Be target
by primary protons were calculated using the 
approach reported in ref. \cite{concha} (see also \cite{van}).\ 
The contribution of protons not interacting 
in the Be target and interacting in the SPS beam dump at the end of the decay
tunnel was also taken into account.\ 
The simulated energy spectra of $\nu_{\tau}$'s pointing to 
the NOMAD fiducial area are shown in Figure 3.\
 The flux of heavy neutrinos, $\Phi(\nus)$, can  then be expressed as follows:

\begin{eqnarray}
\Phi(\nus)\propto \int d\sigma(p + N\rightarrow D_s (\overline{D}_s) + X)
/dE_{\nu_{\tau}}\cdot
Br(D_s(\overline{D}_s)\rightarrow\tau + \nu_{\tau}) \cdot \\ \nonumber
\cdot \tilde{\lambda}^{1/2}\cdot\tilde{h}^{1/2}\cdot|U_{\tau 4}|^2
\cdot dE_{\nu_{\tau}}  
\end{eqnarray}

where $\sigma(p + N\rightarrow D_s (\overline{D}_s) + X)$
is the $D_s$ meson production cross-section \cite{concha,van},
$Br(D_s\rightarrow\tau + \nu_{\tau})$ is 
 the $\tau$-decay mode branching ratio of the $D_s$ \cite{pdg}, and
$\tilde{\lambda}^{1/2},~\tilde{h}^{1/2}$ are the decay phase space 
and helicity factors, respectively \cite{shrock}.\

Once the $\nus$ flux was known, the next step
was to calculate the $\pair$ spectrum based on  
the  differential $\nudecay$ decay 
rate  including  polarization effects (see Figure 4)  \cite{memo}.\ 
The decay electrons and positrons were tracked through 
the DC target including bremsstrahlung photons, their conversion and 
multiple scattering in the target.\ The details 
of the NOMAD simulation and reconstruction are described elsewhere 
\cite{nomad1}.\ 
%The $\ee$ reconstruction efficiency  was computed and convoluted 
%with the DC target details and detector geometrical acceptance based on
%the standard NOMAD MC used in the our previous searches \cite{nomad1}
% and corrected for data themself (see Section 4).\

%% file: analysis
\section{Data analysis and selection criteria}

The search for $\nudecay$ described in this paper uses the full data 
sample collected with the $\overline{V} T_1 T_2$ trigger \cite{nomad} 
during the years 1996-1998.\ The data correspond to a total  
number of protons on target ({\em pots}) of $4.1\times 10^{19}$.\  
The strategy of the analysis was to identify $\nudecay$ candidates 
by reconstructing in the DC isolated low invariant mass $\pair$ pairs 
 that are accompanied 
by no other activity in the detector.\ The measured rate 
of $\pair$ pairs was then compared to that expected from known sources.\

%Candidate events are expected to consist of 
%an $\pair$ pair unaccompanied by any other particles.\
The following selection criteria were applied:

\begin{itemize}
\item two and only two tracks forming a vertex within the DC fiducial volume
of $2.4\times2.4\times3.5 ~\rm m^3$ equivalent to a mass of 1.97 tons; 

\item at least one of the two tracks   identified  as an electron  
 by the TRD  (pion contamination probability $<10^{-3}$ \cite{trd});

\item any additional track or converted photon \cite{profile} 
in the event were allowed only if 
their energies were less than 0.4 GeV or if they could be identified as 
due to  bremsstrahlung photons from one of the two electron candidates;

\item  no $\gamma$'s in the
 ECAL with energy $E_{\gamma}>0.4 ~\rm GeV$ (or $E_{\gamma}>0.3 ~\rm GeV$ for 
$\gamma$'s converted in the PRS), which are incompatible 
with bremsstrahlung photons from the initial tracks;

\item total HCAL energy $<0.4~\rm GeV$.\ This cut serves as an HCAL 
veto and is confirmed by random trigger events and Monte Carlo (MC) studies
\cite{trigger};
 
\item the total energy of the pair must be greater than 4 GeV and its
invariant mass $m_{\pair} < 95~\rm MeV$ to remove background from 
pairs of particles other than $\pair$.

%\item MUON VETO and no FCAL activity, $E_{FCAL}<1~GeV$.
\end{itemize}
 
Only 207 events passed these criteria.\ At the next step we used a 
collinearity  variable ${\mathcal C} \equiv 1- cos \Theta_{\nu \pair}$, where 
$\Theta_{\nu \pair}$ is the angle between the average neutrino beam direction 
and the
total momentum of the reconstructed $\pair$ pair.\ A cut on this variable 
allowed a more effective background suppression.\ 
A MC simulation of heavy neutrino decays shows (see Figure 5) that
 the $\nudecay$ events have  $\C < 2\times10^{-5}$.\
This was true for a $\nus$ mass up to $\simeq 190~\rm MeV$.\ 

In order to avoid biases in the determination of selection criteria, a
blind analysis was performed.\  Events in a signal box 
defined  by  $\C < 2\times10^{-5}$ were excluded from the analysis of the data
until  the validity of the 
background estimate in this region was established.\ This was done by
verifying that the MC simulation of standard processes reproduced the data 
outside the box.\

The  accuracy of the collinearity 
determination obtained with MC simulations was checked using 
a $\nu_{\mu}CC$ data sample with an $\ee$ pair from a photon 
converted  in the DC target at a large ( $\gtrsim 100~\rm cm$) 
distance from the primary vertex.\
Figure 6  shows the  $(1-cos\Theta_{\pair})$ distribution of such events 
in the data and simulation, where $\Theta_{\pair}$ is the  
angle between the $\ee$ pair momentum and the line joining the primary
 vertex to the conversion point.\ The small difference between
the MC and data distributions in Figure 6 would result in an
overall efficiency correction of less than 6\%.\ However, in order to 
conservatively account for possible instrumental effects not present in the MC,
the MC efficiency was multiplied by the efficiency of reconstructing 
$\pair$ pairs with a collinearity variable $\C < 2\times10^{-5}$ in the data
sample of Figure 6 ($\simeq 75\%$).\ Nonetheless,  the two distributions are 
in reasonable agreement at all energies studied.\
This validates the resolution in the variable $\C$ (a few mrad in 
$\Theta_{\nu \pair}$) predicted by the MC program.\

The reconstruction efficiency for the $\nudecay$ decay in the NOMAD fiducial
 volume  was calculated from the MC simulation 
as a function of $\pair$ energy in the 
range  4 GeV to 50 GeV.\ The MC simulation was used to  
correct the data for acceptance losses, experimental resolution and 
reconstruction efficiencies.\ Two checks using both experimental data 
and the MC simulation have been performed in order to verify
the reliability of the simulation and to estimate the systematic uncertainties
in the $\pair$ pair efficiency reconstruction 
in the energy range predicted by the simulation.\\

%Two methods were used in order to check the efficiency of $\ee$ pairs
%reconstruction in NOMAD as a function of energy.\ Reconstructed 
%electron-positron pairs were used to compute the efficiency for signal 
%pairs in the energy range predicted by the simulations, 

The first method is to select two samples of reconstructed $\pi^0$'s, 
one in which the two decay photons reach the ECAL, $N^{\pi^0}_{2\gamma}$, 
and another in which one of the photons converts in the drift chambers,
$N^{\pi^0}_{\gamma\ee}$.\ For data and MC events the ratio

\begin{equation}
R_{Data, MC} =\big( \frac{N^{\pi^0}_{\gamma\ee}}
{N^{\pi^0}_{2\gamma}}\big)_{Data, MC}   
\end{equation}

was then formed.\ The value of the double ratio $RR=R_{Data}/R_{MC}$ 
is then a measure of any differences in $\pair$ reconstruction efficiency
between the data and MC.\ The use of $\pair$ pairs  from 
 $\pi^0$ decay enhances the purity of the $\pair$ sample.\

The method works well mostly for the low 
energy region, $E_{\ee}\lesssim 10~\rm GeV$, 
when the $2\gamma$ opening angle is relatively large 
and the distance between the photons in the ECAL is larger than the
ECAL cell size.\ At higher energies the precision 
of this method is affected by the statistical uncertainties in the number of 
$\pi^0$'s reconstructed 
in the $2\gamma$ mode, because the resolution on 
the $2\gamma$ opening angle becomes worse  
and the $\pi^0$ peak is not well identified anymore.\ 
 
 A similar method allowing a more accurate evaluation of the 
$\ee$ efficiency correction factor at higher energies  
is based on the inclusive $\ee/\gamma$ double ratio
$RR$  defined again as $RR=R_{Data}/R_{MC}$ with     
\begin{equation}
R_{Data, MC}=\big(\frac{N^{\gamma}_{\ee}}{N^{\gamma}_{\gamma}}\big)_{Data, MC}
\end{equation}
where $ N^{\gamma}_{\gamma}, ~N^{\gamma}_{\ee}$ are the numbers of 
single isolated photons and $\ee$ pairs in the same data sample of 
$\nu_{\mu}CC$ events used for the collinearity check.\

It was found that the two methods agree quite well in the low energy region
and yield a correction factor close to 1.\ However, 
in the high energy region the $\ee$ efficiency correction factor 
varied from $0.7\pm0.04$ to $0.4\pm0.03$
 depending on the $\pair$ energy.\

%% file: background
\section{Background}

 The largest contribution to the background is expected  from
 neutrino interactions yielding a single 
$\pi^0$  with little hadronic activity in the final state.\
 Neutrino interactions occuring in the coil and iron upstream 
 of the DC fiducial volume may yield an isolated $\pair$ pair if
 a photon from a $\pi^0$ produced in such an interaction
  converts in the DC  and the  
accompanying particles are not detected.\

Because of the large mass of this upstream material 
the study of this background would require the simulation of 
a very large number of events resulting in a prohibitively large amount of 
computer time.\
Consequently, only about 10\% of the required statistics 
for $\nu_{\mu}CC(NC)$ inelastic reactions were simulated, while other
background components, such as $\nu_e CC$, coherent $\pi^0$ production, 
quasi-elastic reactions and $\nu_{\mu}e$ scattering  were  simulated 
with a statistics comparable to the  number of events expected from  these 
reactions in the data.\ 

%For this reason  the number of background events were
% estimated mainly from the data itself.\ 
%To minimize this background we require  that the cosine of the angle 
%between neutrino direction and total momentum of the $\ee$ pair be smaller 
%than 2$\times 10^{-5}$.\  

The distribution of the variable $\C$ for the sum
of all the MC samples is shown in Figure 5.\
The plot covers the region $\C < 5\times10^{-4}$,
which is 25 times larger than the size of the signal box.\
No  $\nu_{\mu}CC(NC)$ event is found in this region.\ The data outside 
the box, also shown in Figure 5,  are consistent with the MC prediction
(19 events observed and $20\pm4$ events predicted).\ 
The estimate of  background from $\nu_{\mu}CC(NC)$ in the signal box 
is based on the observation that there
are {\em no physical reasons} for this background
 to be other than flat in the region $\C < 5 \times 10^{-4}$.\
% except for $\nu_{\mu} e$-scattering which was found to be small.\ 
Two independent methods were used  for the 
background estimation in the signal region.\ 

The first method is based on the MC.\
The  MC background from the fully simulated reactions
 was found to be $2.5\pm 0.8$ events inside the signal region.\
For the $\nu_{\mu} CC$ and $\nu_{\mu} NC$ events 
for which only 10 \% of the data statistics was generated, no event  
was found in the full enlarged
region $\C < 5\times 10^{-4}$.\ 
Assuming that this background is distributed randomly in this interval and 
taking into account that the simulated sample corresponds to only 10\% of 
data, we estimate the $\nu_{\mu} CC$ and $\nu_{\mu} NC$ background 
contribution to be  $0^{+0.4}_{-0.0}$ events inside the signal region.\

The second method relies on the data themselves.\ 
% The total number of observed events in the enlarged region
%$\C < 5\times 10^{-4}$ outside the signal box is 19.\ The corresponding
% background from the fully simulated reactions is $20.1\pm 4$ events.\
The agreement between the observed and predicted  numbers
 of events outside the signal box 
allows to conclude that the number of background events from 
$\nu_{\mu}CC$ and $\nu_{\mu}NC$ processes is negligible.\
Thus, by extrapolating the 19 observed events to the signal region with 
the shape of the fully simulated MC events we obtain a second 
background estimate of $2.4\pm0.9$.

 The final background  estimates with the two methods are 
$N_{bkg}^{\rm MC}=2.5^{+0.9}_{-0.8}(stat)$ $\pm 0.6 (syst)$ events from the MC 
and $N_{bkg}^{\rm Data}= 2.4^{+0.9}_{-0.9}(stat) \pm 0.7 (syst)$ events
from the data, thus providing consistent results.\
 The systematic error includes
the uncertainties in the number of {\em pot} (5\%) and in the coherent
 $\pi^0$ production cross section (25\%).\ In addition,
in the second method we also take into account the systematic errors
related to the extrapolation procedure.\
The total systematic uncertainty  was calculated by adding all errors in 
quadrature.\ 
% and decay branching ratios (20\%) and 
%$\ee$ pair reconstruction efficiency (10\%).\ 
In the following we use the background estimate 
extracted from the data themselves.\

%% file: result_p
\section{Results and calculation of limits} 

Upon opening the signal box we have found
one event that passes our selection criteria.\ 
This is consistent with the expected background and hence 
no evidence for isosinglet neutrino decays has been found.\ 
We can then determine the $90\%~ CL$
upper limit for the corresponding mixing amplitude $|U_{\tau 4}|^2$
from the $90\%~ CL$ upper limit for the expected number of signal events,
$N_{\nudecay}^{\rm up}$.\ Using the frequentist approach of 
ref. \cite{gary} and  taking into account the uncertainties in the background 
estimate \cite{stuart} we obtain $N_{\nudecay}^{\rm up}=2.1$ events. 
Since the number of observed events is less than the predicted background, 
this limit is lower than the corresponding sensitivity, 
$S_{\nudecay}^{\rm up}=4.0$. The sensitivity is defined as the average 
upper limit obtained in the absence of signal events by 
an ensemble of experiments with the same expected background \cite{gary}.
The probability to obtain an upper limit of 2.1 or lower is 29\%.  

For a given flux  $\Phi(\nus)$, the expected number of 
$\nudecay$ decays occuring within the fiducial length $L$ of the NOMAD 
detector located  at a distance $L'$ from the
neutrino target is given by 

\begin{eqnarray}
N_{\nudecay} =\int \Phi(\nus)\cdot exp(-L'm_{\nus}/p_{\nus}\tau_4)\cdot 
[1-exp(-L m_{\nus}/p_{\nus}\tau_4)] \\ \nonumber
\cdot (\Gamma_{\pair}/\Gamma_{tot})\cdot \varepsilon\cdot A \cdot dE_{\nus}\propto |U_{\tau 4}|^4
\end{eqnarray}

where $p_{\nus}$ is the $\nus$ momentum and  $\tau_4$ is its 
lifetime at rest, $\Gamma_{\pair},~\Gamma_{tot}$
are the  partial and total  mass dependent 
$\nus$-decay widths, respectively, and 
$\varepsilon$ is the $\pair$ pair reconstruction efficiency.\ The acceptance 
$A$ of the NOMAD detector was calculated  tracing $\nus$'s
produced in the Be-target or beam dump to the detector taking all 
relevant momentum and angular distributions into account \cite{concha}.\ 
As an example for a mass $m_{\nus}=33.9~\rm MeV$, $A=1.4\%$ and $\varepsilon=26\%$.\
The flux $\Phi(\nus)$ is given by  Eq. (4).\ 
%Over most of the $\tau_{\nu_s}$ region, the $\nu_s$ neutrino 
%lifetime is sufficiently long that 
%\begin{equation}
%Lm_{\nu_s}/p_{\nu_s}\tau \ll L'm_{\nu_s}/p_{\nu_s}\tau \ll 1
%\end{equation}

%Thus we may write,

%\begin{equation}
%N_{\nudecay}=\int \Phi(\nu_s)\cdot (Lm_{\nu_s}/p_{\nu_s}\tau_p)\cdot (\Gamma_{\pair}/\Gamma_{tot})\cdot \varepsilon\cdot A \cdot dE_{\nu_s}\propto |U_{\tau 4}|^4
%\end{equation}

%When the $\nu_s$ lifetime is very short such that 
%$L'm_{\nu_s}/p_{\nu_s}\tau >1$,
%the approximation of Eq.(1) is no longer valid.\ However, as one can 
%easily see
%the lower value for the KARMEN sterile neutrino lifetime is of the order of 
%$10^{-1}\mu s$ (see Fig.1).\ For this lifetime the $\nu_s$-decay 
%length is of the 
%order of $L_{decay} = c \gamma \tau \approx 24~km$
% for $\gamma\simeq <E_{\nu}>/m$, where $<E_{\nu}>\approx 26$ GeV, and 
%$m = $33.9 MeV/$c^2$.\ Thus the approximation of the Eq.(1) is valid.
 
The final $90\%~ CL$ upper limit curve  in the 
($m_{\nus}; |U_{\tau 4}|^2$) plane 
is shown in Figure 7 together with the Big Bang Nucleosynthesis (BBN) and 
Supernova SN1987a lower limits obtained in 
ref. \cite{dolgov1}.\ For the mass range $m_{\nus}\gtrsim m_{\pi}$ the 
shape of the NOMAD limit curve is explained by the contribution from the
$\nus \to \nu_{\tau}\pi^0$ decay  to the $\nus$
decay rate: as the $\nus \to \nu_{\tau}\pi^0$ decay channel opens up, 
the branching ratio BR($\nudecay$)  drops rapidly    
as  $m_{\nus}$ increases.\ 

%The $90\%~ C.L.$ exclusion plot in the ($m_{\nu_s}; |U_{\tau 4}|^2$) plane 
%is shown in Figure 1.\ The plot was calculated summing up the contributions
%from both $\nu_s$ and $\overline{\nu}_s$ for Dirac case and from $\nu_i$
%coming from both $D_s(\overline{D}_s)$ decays for Majorana case, since in 
%both cases the $\ee$ pairs in the final state are undistinguishable.\

% Contribution of the systematic errors to the NOMAD limit was estimated 
%based on the following procedures.\ 

%% file: karmen1
\section{The KARMEN time anomaly}

Our data  can also be used to restrict one of the interpretations 
of the time anomaly observed by the KARMEN experiment \cite{karmen}.\
The anomaly is a bump in the time distribution 
of $\nu_e$ and $\overline{\nu}_{\mu}$ induced 
events which was expected to be well
described by the single exponential from  muon decays.\footnote{This
 anomaly, seen in the KARMEN1 data, was not confirmed with the KARMEN2
data \cite{eitel}.\ A possible 
explanation of this effect may be found in ref. \cite{atch}.} 
 The KARMEN Collaboration interpreted 
the anomaly as being due to an exotic decay of  $\pi^+$-mesons into a muon 
and a
33.9 MeV new fermion $X$ and reported a signal curve for pion 
branching ratio $BR(\pi^+\to \mu^+ + X) \times \Gamma (X\to \pair \nu)$ as a 
function of $X$-lifetime with a  branching ratio
$BR(\pi^+\to \mu^+ + X)$ as small as $10^{-16}$ \cite{karmen}.\ 
This and other hypotheses explaining the anomaly
 have been recently extensively investigated both
theoretically \cite{barger,dolgov2,theory1} and 
experimentally \cite{daum,exper}.\

Barger et al. \cite{barger}  
associated the new particle with a 33.9 MeV isosinglet neutrino 
dominantly mixing with $\nu_{\tau}$.\  From our limit curve in Figure 7,
we have found using Eqs. (2,3) that for $m_{\nus}=33.9~\rm MeV$ the lifetime 
of such a  neutrino has to be greater than $\simeq 10^{-2}$ sec while the 
BBN lower limit of ref. \cite{dolgov2}
constrain the $\nus$ lifetime to be less than 0.1 sec.\footnote{The SN1987a
limit obtained is stronger than the BBN one, however the 
supernova results are more model dependent.\ One may imagine a case when the
supernova arguments are not applicable, while the BBN ones still 
apply \cite{dolgov2}.}  
%Considering the $\pi^+\to \mu^+ + X$ decays of 
%secondary pions in the WANF decay 
%tunnel as the source of $X$'s one can calculate the excess of $\pair$ pairs
%from $X\to \pair \nu_4$ decays in NOMAD as a function of 
%$BR(\pi^+\to \mu^+ + X)$ \cite{memo}.\ No excess was found.\  
Using the KARMEN signal curve \cite{karmen} results in a 
small window around $BR(\pi^+\to \mu^+ + X)\simeq 10^{-12}$ between our 
result \cite{memo} and BBN one left untested.\ 
 Note that the recent PSI result  on a search for the
33.9 MeV particle in $\pi^+\to \mu^+ + X$ decay corresponds to 
$BR(\pi^+\to \mu^+ + X)<6.0\cdot 10^{-10}$ at 95\% $CL$ \cite{daum}.\

%% file: conclusion
\section{Conclusion}

 We found no evidence for the existence of a 
heavy neutrino, $\nu_4$, with  mass in the 
range 10 - 190 MeV which  mixes dominantly with the 
$\nu_{\tau}$  and decays into $\nu_{\tau} e^+ e^-$.\   
For the first time an upper limit on the square of the  
mixing amplitude, $|U_{\tau 4}|^2$, is obtained.\
Windows between our result and the BBN and SN1987a limits are still of interest
for further searches.

\vspace{.5 cm}

{\large \bf Acknowledgements}

\vspace{.5cm}
We gratefully acknowledge  the CERN SPS accelerator and beam-line staff
for the magnificent performance of the neutrino beam.\
The experiment was  supported by  the following
funding agencies:
Australian Research Council (ARC) and Department of Industry, Science, and
Resources (DISR), Australia;
Institut National de Physique Nucl\'eaire et Physique des Particules (IN2P3), 
Commissariat \`a l'Energie Atomique (CEA),  France;
Bundesministerium f\"ur Bildung und Forschung (BMBF, contract 05 6DO52), 
Germany; 
Istituto Nazionale di Fisica Nucleare (INFN), Italy;
Russian Foundation for Basic Research, % (grant 96-02-18562), 
Institute for Nuclear Research of the Russian Academy of Sciences, Russia; 
Fonds National Suisse de la Recherche Scientifique, Switzerland;
Department of Energy, National Science Foundation (grant PHY-9526278), 
the Sloan and the Cottrell Foundations, USA. F.J.P. Soler is
supported by a TMR Fellowship from the European Commission.\
M.C. Gonzalez-Garcia is supported by the Spanish DGICYT under grants 
PB98-0693 and PB97-1261, by the Generalitat Valenciana under grant
GV99-3-1-01 and by the TMR network grant ERBFMRXCT960090 of the
European Union.

We thank Prof. V.Matveev  for many interesting discussions and 
comments.\ 
We should like to thank Prof. A.D. Dolgov  and Dr. D.V. Semikoz for their 
interest in our results and for many interesting 
discussions on the astrophysical and cosmological limits.